\documentclass[twocolumn,amsmath,amssymb]{revtex4}

\usepackage[dvips]{graphicx}
\usepackage{amssymb,amsfonts,amsmath,subfigure}

\usepackage[latin1]{inputenc}

\usepackage{indentfirst}

\usepackage{amsfonts}
\usepackage[normalem]{ulem}

\usepackage{latexsym}
\usepackage{amssymb}
\usepackage{amsmath}
\usepackage[dvips]{graphicx}

\usepackage{float,epsfig,color,array}

\def\1{\mathds{1}}

\newcommand{\N}{\mathbb N}

\newcommand{\E}{\mathbb E}

\newcommand{\diag}{\mathop{\mathrm{diag}}\nolimits}
\newcommand{\trace}{\mathop{\mathrm{trace}}\nolimits}

\begin{document}

\title{Stability of graph communities across time scales}
\date{\today}

\author{J.-C. Delvenne$^{1}$\footnote{Present address:
Department of Applied Mathematics, Universit\'e catholique de
Louvain, , 4, av. Lema\^itre, B-1348 Louvain-la-Neuve, Belgium},
S.N. Yaliraki$^{1,2}$, and M. Barahona$^{1,3}$}

\address{$^1$Institute for Mathematical Sciences,$^2$Department of Chemistry,$^3$Department of Bioengineering\\
Imperial College London, South Kensington Campus, London SW7 2AZ,
United Kingdom}

\begin{abstract}
The complexity of biological, social and engineering networks makes
it desirable to find natural partitions into communities that can
act as simplified descriptions and provide insight into the
structure and function of the overall system. Although community
detection methods abound, there is a lack of consensus on how to
quantify and rank the quality of partitions.  We show here that the
quality of a partition can be measured in terms of its stability,
defined in terms of the clustered autocovariance of a Markov process
taking place on the graph. Because the stability has an intrinsic
dependence on time scales of the graph, it allows us to compare and
rank partitions at each time and also to establish the time spans
over which partitions are optimal. Hence the Markov time acts
effectively as an intrinsic resolution parameter that establishes a
hierarchy of increasingly coarser clusterings. Within our framework
we can then provide a unifying view of several standard partitioning
measures: modularity and normalized cut size can be interpreted as
one-step time measures, whereas Fiedler's spectral clustering
emerges at long times.  We apply our method to characterize the
relevance and persistence of partitions over time for constructive
and real networks, including hierarchical graphs and social
networks.  We also obtain reduced descriptions for atomic level
protein structures over different time scales.
\end{abstract}

\maketitle

\section{Introduction}

In recent years, there has been an explosion of interest in the
analysis of networks as models of complex systems. The literature is
extensive spanning areas as diverse as gene regulation, protein
interactions and metabolic pathways, neural science, social networks
or engineering systems such as transportation networks and the
internet, to name but a few~\cite{Strogatz01, Newman03}. The tools
for network analysis are firmly grounded on results in graph theory,
with an influx of concepts from statistical physics, dynamical
systems and stochastic
processes~\cite{DaCostaRodriguesTraviesoVillasBoas07}. Due to the
large-scale, complex nature of many systems under study, an
appealing idea is to obtain relevant partitions (or clusterings) of
the network that can reveal the underlying structure of the system
and hence insight into its function. These partitions could
potentially lead to reduced, more manageable representations of the
original system.

The topic of graph community detection has a long history and
multiple methods and heuristics have been proposed to partition
graphs into communities or clusters. (See for
instance~\cite{fortunato_survey} and references therein for a recent
survey.) However, the extensive list of partitioning methods comes
with a parallel lack of theory or consensus on measures to quantify
the goodness of a community structure. The simplest such measure is
certainly the \textit{cut size}, i.e., the sum of the weights of
edges that lie at the boundaries of different communities. As a
general rule, good community structures should have small cut size
implying that the communities are weakly connected. Unfortunately,
this simple intuitive notion has negligible applicability since the
partition with minimum cut size is often trivial. Therefore, a
variety of measures have been proposed including, without claim of
exhaustivity, normalized cut~\cite{ShiMalik00},
$(\alpha,\epsilon)$-clustering~\cite{Kannan07},
modularity~\cite{NewmanGirvan04,Newman06PNAS} and variants and
extensions of
modularity~\cite{MuffRaoCaflisch05,ReichardtBornholdt06}. Each of
these methods has distinct features and has been shown to produce
reasonable community structures for different examples. In
particular, modularity does not require that the number of
communities be specified in advance, unlike most of the other
partitioning methods. However, it has been recently shown that
optimizing modularity can over-partition or under-partition the
network, failing to find the most natural community
structure~\cite{FortunatoBarthelemy07}. To compensate for this,
recent
methods~\cite{ReichardtBornholdt06,ArenasFernandezGomez07,LancichinettiFortunatoKertesz08},
have included an \textit{ad hoc} resolution parameter that can be
tuned to bias towards small or large communities. The introduction
of these resolution parameters highlights the fact that one would
expect that any given graph would be described by different natural
community structures (finer or coarser) under different conditions.

Our work introduces a quality measure that has the intrinsic
flexibility to find which clusterings are relevant at different time
scales. This is achieved by establishing a link between the quality
of the partition and a stochastic process taking place on the
clustered graph. We use the well-known relationship between graphs
and Markov chains: with any unweighted graph we can associate a
random walk in which the probability of leaving a vertex is
uniformly distributed among the outgoing edges. This Markov
viewpoint provides a dynamical interpretation of communities. In
particular, natural communities at a given time scale will
correspond to persistent dynamical basins, that is, sets of states
from which escape is unlikely within the given time scale. This can
be established quantitatively through the autocovariance of the
clustered Markov process, a measure that defines the persistence of
a cluster in time. In essence, one can think of the time scale as an
intrinsic \textit{resolution parameter} for the clustering:  over
short time scales, many clusters should be coherent; on the other
hand, the expectation is that there will be few persistent clusters
under the action of the Markov chain  if one waits for a long time.

An important feature of our approach is that it provides a framework
that unifies several heuristic measures. It turns out that most
quality measures introduced in the literature have a natural Markov
probabilistic interpretation. We will show below that modularity and
normalized min-cut are related to the autocovariance on paths of
length one (i.e., at time $t=1$), while Fiedler's spectral method
corresponds to the limit of long paths (i.e, time $t=\infty$). In
contrast, our measure considers paths of all lengths and provides an
evaluation of the quality of a clustering at all times, including
fractional times ($0<t<1$) for which we obtain clusterings finer
than those obtained by modularity optimization. Our measure is thus
not affected by the resolution limit of
modularity~\cite{LancichinettiFortunatoKertesz08}.

The rest of the paper proceeds by introducing in simple terms the
definition of the \textit{stability of the clustering} $r(t)$, which
corresponds to the autocovariance of the partition under a Markov
process and provides a measure of the quality of any partition over
time. As part of our derivation, we show that $r(1)$ is optimized by
modularity while at long time scales, $r(\infty)$ is typically
optimal for the classic 2-way spectral clustering related to the
Fiedler vector. For the intermediate time scales, our measure can be
used to rank the different partitions and, in doing so, establish a
hierarchical,  time-dependent set of partitions that are valid over
different time spans.  Our measure also allows us to compare the
community structure obtained by different algorithms over different
timescales. In addition, we show how the stability at fractional
times,  $r(0<t<1)$, leads to finer partitions than those produced by
modularity maximization. Therefore the stability $r(t)$ provides a
unifying framework for the understanding of different and seemingly
unrelated clustering heuristics in relation to the characteristic
Markov time over which a given clustering is valid. We exemplify the
applications of the method with networks drawn from different fields
to showcase the generality of the approach.

\section{Methods}

\subsection{Autocovariance and stability of a graph partition}

Consider an undirected, connected graph with $N$ vertices and $E$
edges and assume that the graph is non-bipartite. For simplicity in
the derivation below, we will assume that the graph is unweighted,
although all our results apply equally to weighted graphs. The
topology of the graph is given by the $N \times N$ adjacency matrix
$A$, a symmetric $0$-$1$ matrix with a $1$ if two vertices are
connected and $0$ otherwise. The number of edges of each vertex, or
degree, $d_i$ can be compiled into the vector $\mathbf{d} = A
\mathbf{1}$, where $\mathbf{1}$ is the vector of ones. We will also
use the diagonal matrix of degrees: $D=\diag(\mathbf{d})$.

A random walk on any such graph defines an associated Markov chain
in which the probability of leaving a vertex is split uniformly
among the outgoing edges, with a transition probability $1/d_i$ for
each edge:
\begin{equation}
\label{eq:Markov} \mathbf{p}_{t+1}=\mathbf{p}_{t}  \left [ D^{-1} A
\right ] \equiv \mathbf{p}_{t}  M,
\end{equation}
where $\mathbf{p}_{t}$ is the (normalized) probability vector and
$M$ is the transition matrix. Under these assumptions, we have an
ergodic and reversible Markov chain with stationary distribution
$\mathbf{\pi}= \mathbf{d}^T/\sum_i d_i = \mathbf{d}^T/2E$. We will
also use below the diagonal matrix $\Pi = \diag(\pi)$.

Consider now a given partition of the graph in $c$ communities. This
(hard) clustering can be encoded in an $N \times c$ indicator matrix
$H$, a 0-1 matrix that records which vertex belongs to which
community. Each row of $H$ is all zeros except for a $1$ indicating
the cluster to which the vertex belongs. Let us now observe the
Markov process~(\ref{eq:Markov}) under the prism of the given
partition by assigning a different real value $\alpha_i$ to the
vertices of each of the $c$ clusters. The observed signal is then a
stationary, not necessarily Markovian, random variable $(X_t)_{t \in
\N}$ which consists of a sequence of  $\alpha_i$. The expectation
for a good partition of the graph over a given time scale is that
the state is likely to remain within the starting cluster for such a
time span, as compared with that event occurring at random. This can
be quantified through the autocovariance of the observable ${\rm
cov}[X_t,X_{t+\tau}] = \E[X_t X_{t+\tau}]-\E[X_t]^2$, where $\E$
denotes expectation. If the inter-community connections are weak,
the values of $X_t$ and $X_{t+\tau}$ will be correlated for longer
times. How fast the autocovariance decays as a function of the lag
$\tau$ is therefore an indicator of the quality of the clustering
over the corresponding Markov time scale. This is the main idea
underpinning our measure.

The covariance of $X_t$ can be rewritten as ${\rm
cov}[X_t,X_{t+\tau}] = \alpha^T R_\tau \alpha$, where $\alpha$ is
the vector of labels of the $c$ communities and the matrix $R_t$ is
the \textit{clustered autocovariance matrix} of the graph:
\begin{equation}
\label{eq:autocovariance} R_t = H^T \left(\Pi M^t -\pi^T \pi \right
) H.
\end{equation}
Note that the matrix $R_t$ depends only on properties of the graph
and clustering. It summarizes the $t$-step dependence of the
transfer probabilities between clusters: each element $\left ( R_t
\right )_{ij}$ corresponds to the probability of starting in a
cluster $i$ and being in another cluster $j$ after $t$ steps minus
the probability that two independent random walkers are in $i$ and
$j$, evaluated at stationarity.

As stated above, a good partition over a given time scale should
imply a high likelihood of remaining within the starting community.
In terms of the clustered autocovariance matrix, the diagonal
elements $\left(R_{t}\right)_{ii}$, which measure the probability of
a random path of length $t$ to start and end in the same community,
should be larger than the off-diagonal ones. This leads to our
definition of the \textit{stability of the clustering}:
\begin{eqnarray}
r(t; H) = \min_{0 \leq s \leq t} \sum_{i=1}^c (R_{s})_{ii} = \min_{0
\leq s \leq t} \trace \left[ R_s \right]. \label{eq:stability}
\end{eqnarray}
A good clustering over time $t$ will have large stability, with a
large trace of $R_t$ over such a time span. Note that our definition
involves the minimum value of the trace in a given interval, i.e.,
the stability of the partition is large only if the values
\textit{for all times} up to $t$ are large. In this way, we assign
low stability to partitions where there is a high probability of
leaving the community and coming back to it later, as in the case of
almost bipartite graphs.

\begin{figure}[h!]
\begin{center}
\includegraphics[width=7.5cm]{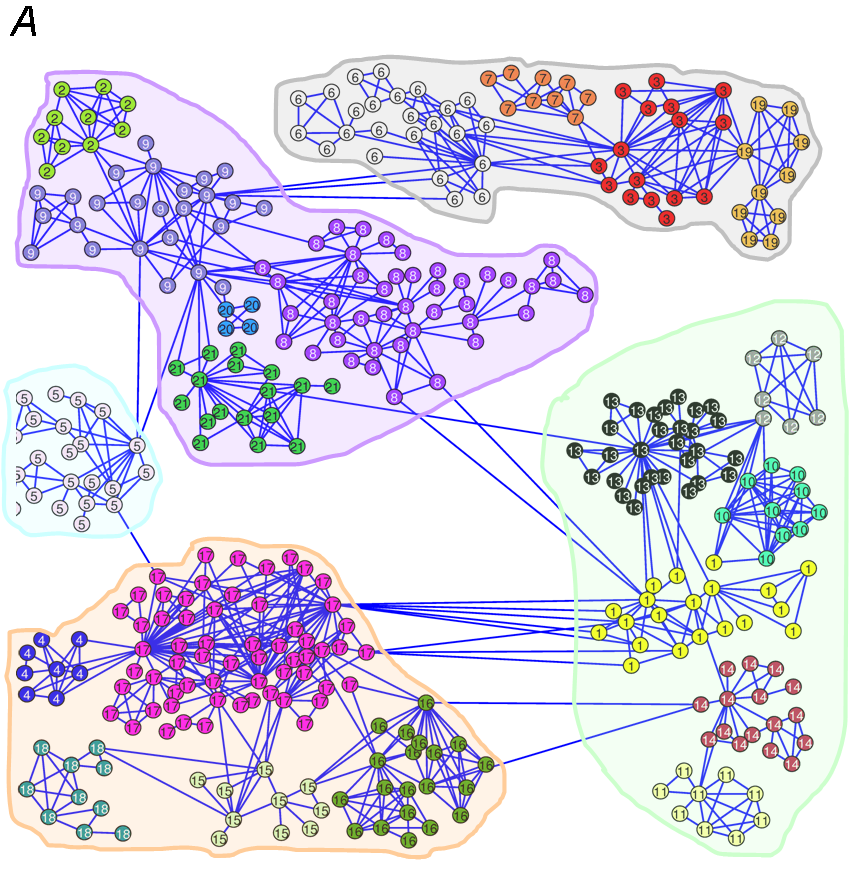}
\\
\includegraphics[width=7.5cm]{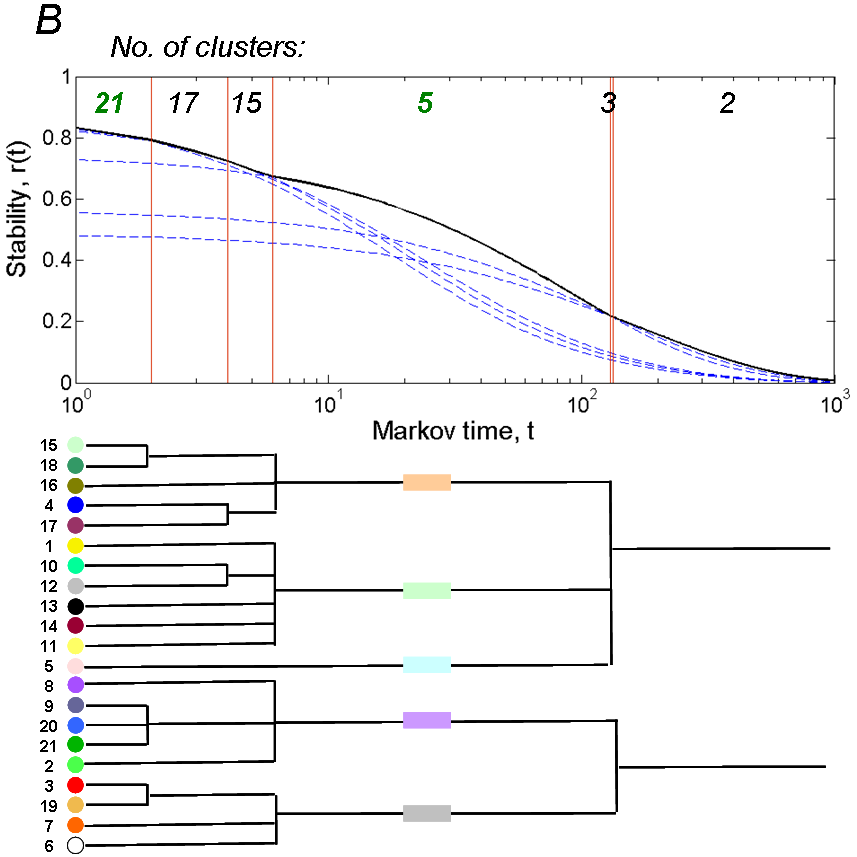}
\end{center}
\caption{\textit{(A)} Largest connected component of a graph of
scientific collaborations in network science~\cite{Newman06}. The
vertices correspond to $N=379$ researchers indexed by the 21-way
partition obtained by maximizing the stability at $t=1$ (or
equivalently, modularity). A list of names for this graph and
groupings is available in the Supplementary Information.
\textit{(B)} Stability curve obtained with the divisive KVV
algorithm (top) and the corresponding dendrogram of the hierarchy of
partitions (bottom). Note the simplicity of the dendrogram, which is
not a binary tree, as compared with the many branching points
obtained by standard binary partition methods. Only two clusterings
are long-lived: the two-way clustering (trivially) and the five-way
partition represented by areas shaded in different colors in
\textit{(A)}. } \label{FigNetSci}
\end{figure}

The stability~(\ref{eq:stability}) is the fundamental tool we
propose to assess the quality of different clusterings over time.
For each candidate clustering, we can compute the stability at all
times and rank the possible partitions. Clearly,  certain partitions
might only be optimal in particular time windows and different
partitions will be optimal at different times. For each Markov time
$t$, we seek the partition with the largest stability to obtain the
\textit{stability curve of the graph}: $r(t) = \max_{H} r(t; H)$.
This curve establishes a time hierarchy of partitions,  from finer
to coarser as time grows, as shown in Figure~\ref{FigNetSci} for a
social network. This underscores the idea that partitions are better
or worse depending on the time of interest, and the concept of the
Markov time as an intrinsic resolution parameter that establishes
when a partition is good. In this sense, the most relevant
partitions will be optimal over long time windows, because they
serve as good representations over extended time scales of the
system.

\subsection{Relationship of the stability with modularity, cut, normalized cut and spectral partitioning}
\label{subsect:relation}

An important feature of the stability~(\ref{eq:stability}) is that
it encompasses several of the criteria for clustering in the
literature and allows us to interpret those heuristics in terms of
the relevant Markov time scales of the graph. To explore this, we
study the autocovariance $R_t$ and the stability $r(t)$ in different
limits.

First, consider short times. At time $t=0$, the partition with the
largest stability is the finest possible clustering. This follows
from the definition $r(0)=  1-\|\pi H\|^2_2$, which becomes maximal
when each vertex is in its own cluster as follows from elementary
inequalities.

At time $t=1$, we recover modularity, a popular measure for
community detection~\cite{NewmanGirvan04}. It follows from the
definition that modularity is equal to the trace of $R_1$, the
autocovariance matrix at time $t=1$.  Therefore, maximizing $r(1)$
is equivalent to modularity optimization. (See also~\cite{Newman06}
for an alternative, non-dynamical take on this issue.) The stability
is also related to other measures in the literature. Consider the
cut size (\textrm{Cut}), defined as the sum of the number of
inter-community edges divided by the total number of edges of the
graph. It is easy to see that $\textrm{Cut}=r(0)-r(1)$. Hence
modularity is equal to $1-\textrm{Cut}-\|\pi H\|_2^2$, and
maximizing modularity is equivalent to minimizing
$\textrm{Cut}+\|\pi H\|_2^2$. This is the reason why modularity
tends to produce balanced partitions: minimizing $\textrm{Cut}$
favors few clusters, possibly of very unequal sizes, while
minimizing $\|\pi H\|_2$ tends to favor many clusters of equal size.
An alternative measure to modularity is the so-called Normalized Cut
size (NCut)~\cite{ShiMalik00}. For the case of two communities, NCut
is the number of inter-community edges  multiplied by the sum of the
inverse of the number of edges in each community, which can be shown
to equal $\textrm{NCut}=\rho(0)-\rho(1)$, where $\rho(t)$ is given
by the same expression as the stability $r(t)$ replacing covariances
by correlations. Hence NCut is also a one-step measure.

The discussion above shows that modularity, Cut and NCut are based
on the one-step behavior of the Markov process. On the other hand,
our stability measure takes into account the dependence of the
autocovariance at all times. In fact, the behavior of $r(t)$ in the
long time limit $t \to \infty$ establishes a link with spectral
clustering methods, the other standard toolbox for graph
partitioning. Spectral methods are generally based on the Fiedler
eigenvector, i.e., the eigenvector associated with the second
smallest eigenvalue of the Laplacian matrix $L=D-A$, or of the
normalized Laplacian  $\mathcal{L}=D^{-1/2} L D^{-1/2}$. In
Fiedler's original work~\cite{Fiedler73,Fiedler75}, the graph was
partitioned into two subgraphs according to the sign of the
components of the Fiedler vector. More recently, graph partitioning
based on the normalized Fiedler vector has been
proposed~\cite{VanDriessche95} and shown to be a heuristic for the
optimal NCut $2$-way clustering~\cite{ShiMalik00}.

The analysis of our measure shows that spectral clustering is not
just a heuristic but an exact method to find the most stable
partitions at long time scales. This follows from the spectral
decomposition of the normalized Laplacian $\mathcal{L}$, which is
trivially related to that of $\mathcal{M}=D^{1/2} M
D^{-1/2}=\sum_{i=1}^N \lambda_i \mathbf{u}_i \mathbf{u}_i^T$. Here
the eigenvalues $\lambda_i$ are ranked in order of decreasing
magnitude and the corresponding eigenvectors $\mathbf{u}_i$ are
orthonormal. In particular, $\lambda_1=1$ and
$\mathbf{u}_1=(1/\sqrt{2E})\, D^{1/2} \mathbf{1}$ leading to the
following asymptotic behavior:
\begin{equation}
\label{eq:spectral_infinity} \trace[R_t] =  \sum_{i=2}^N
\frac{\lambda_i^t}{2E} \|H^T D^{1/2} \mathbf{u}_i \|^2 \xrightarrow
{t \to \infty} \frac{\lambda_2^t}{2E} \|H^T D^{1/2} \mathbf{u}_2
\|^2.
\end{equation}
If $\lambda_2$ is positive, $\mathbf{u}_2$ is the normalized Fiedler
eigenvector and the clustering with maximal stability at long times
typically corresponds to the Fiedler partition. To see this, take
initially the finest possible partition with each node in a cluster
by itself and cluster together vertices $i$ and $j$. This induces a
variation in~(\ref{eq:spectral_infinity}) given by $(\lambda_2^t/E)
\sqrt{d_i d_j} \, u_{2,i} u_{2,j}$, which is only positive if the
components of the normalized Fiedler vector for nodes $i$ and $j$
have the same sign. Applied recursively, this leads to the result
that the partition with the largest stability at long times is
typically the 2-way clustering according to the sign of the entries
of the Fiedler vector.

When $\lambda_2$ is negative, $\mathbf{u}_2$ is not the Fiedler
eigenvector but rather the largest eigenvalue of $\mathcal{L}$,
i.e., the most negative eigenvector of $\mathcal{M}$. In this case,
the dominant term in~(\ref{eq:spectral_infinity}), and hence the
stability, becomes negative for all partitions except for the
coarsest clustering with all nodes in one community and $H =
\mathbf{1}$, for which the stability is zero at all times, following
from~(\ref{eq:spectral_infinity}) and orthogonality. We thus
conclude that, at large times, the clustering with maximal stability
is either a one-way or two-way partition.  In the latter case, it is
given by the normalized Fiedler vector.

The overall picture emanating from our analysis is that the
partition with highest stability evolves from the finest possible
(each vertex by itself) at $t=0$, through the optimal modularity
clustering at $t=1$, onto a sequence of  coarser partitions, the
last of which is typically the two-way spectral clustering (or the
one-way trivial clustering) as $t \to \infty$. Although the sequence
of partitions is not necessarily always increasingly coarser at
increasing times (we may have incomparable clusterings that are
optimal at different times), we do expect  that the clusterings will
roughly contain fewer and fewer clusters as the Markov time grows.

\section{Applications and examples}
\label{sect_appl}

We now show the applicability of the method by analyzing three
examples drawn from social interactions, hierarchical scale-free
graphs and protein structural networks. Rather than being
exhaustive, our goal is to highlight through each example some of
the wider features of our approach.

\subsection{Example 1 -- Time hierarchy of partitions and comparison of clustering algorithms}

Our first example deals with the graph of collaborations between
researchers in network science shown in
Figure~\ref{FigNetSci}\textit{A}~\cite{Newman06}. Community
structures are relevant for social networks, where the
identification of groups of people with strong ties can help unravel
underlying patterns of
interdependence~\cite{DaCostaRodriguesTraviesoVillasBoas07}. In
Figure~\ref{FigNetSci}\textit{B} we show the time hierarchy of
partitions associated with the stability curve of the network. Our
measure~(\ref{eq:stability}) is used to rank partitions efficiently,
since the stability of a given clustering $r(t; H)$ is directly
computable in $\mathcal{O}(cEt)$, or estimated in $\mathcal{O}(Kt)$
with accuracy $\mathcal{O}(c/\sqrt{K})$ through $K$ random walks of
length $t$. In order to obtain the stability curve, one needs to
maximize the stability over all partitions. Given that modularity
optimization is provably NP-hard~\cite{Brandes07}, it is likely that
no efficient algorithm exists for the optimization of stability for
arbitrary graphs. However, for all practical applications, we can
still obtain sequences of partitions through the use of a number of
partitioning algorithms with different heuristic strategies, such as
aggregative (i.e., unifying clusters from the finest clustering) or
divisive (i.e., splitting clusters from the coarsest clustering).
Figure~\ref{FigNetSci}\textit{B} is the result of the application of
Kannan, Vempala and Vetta's (KVV) conductance spectral
algorithm~\cite{Kannan07} under a divisive strategy to produce a
sequence of partitions, which are then ranked according to their
stability to estimate the stability curve $r(t)$. This curve is then
translated into a non-binary dendrogram representing the sequence of
community structures with maximal stability as a function of time.
The dendrogram has the advantage of being relatively simple, with
fewer branching points compared with the binary trees produced by
most hierarchical community detection algorithms. In this case, the
time hierarchy of partitions indicates that the modularity-optimal
clustering into $21$ communities is short-lived whereas a partition
into $5$ communities persists over a long time window. This suggests
the relevance of this coarser meta-community structure as indicative
of the likelihood of information to flow within the five subgroups
of researchers.

\begin{figure}[t]
\begin{center}
\includegraphics[width=8cm]{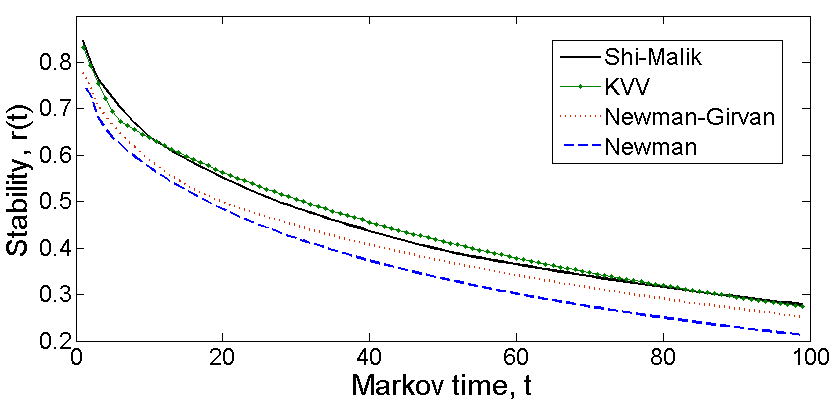}
\end{center}
\caption{A comparison of the stability curve of the partitions
obtained through a divisive strategy using four clustering
algorithms (Shi-Malik~\cite{ShiMalik00}, KVV~\cite{Kannan07},
Newman~\cite{Newman06} and Newman-Girvan~\cite{NewmanGirvan04}) on
the network of scientific collaborations pictured in
Fig.~\ref{FigNetSci}\textit{(B)}} \label{fig:fouralgos}
\end{figure}

Our stability measure can also be used to rank the sequences of
partitions obtained by different algorithms and strategies.
Figure~\ref{fig:fouralgos} presents the comparison of the estimated
stability curves from four algorithms chosen for their simplicity
and popularity and because they represent different overall
methodologies. In addition to the KVV conductance method introduced
above~\cite{Kannan07}, we have also examined Shi-Malik's recursive
spectral method~\cite{ShiMalik00}, Newman's spectral method to
optimize modularity~\cite{Newman06} and the Newman-Girvan
betweenness algorithm \cite{NewmanGirvan04}. In all cases, we use a
divisive strategy to produce a sequence of increasingly finer
$k$-way partitions and obtain an estimate of the stability curve
$r(t)$ by choosing the best partition at each time. For details of
the algorithms see the Supplementary Information.
Figure~\ref{fig:fouralgos} shows that Shi-Malik and KVV produce the
partitions with highest stability at all shown times (alternatively
better in different time windows), followed closely by the
Newman-Girvan algorithm and Newman's spectral algorithm. At higher
times (up to $t=1000$ at least), the KVV method slightly dominates
Shi-Malik and Newman-Girvan algorithms, while Newman's clustering
algorithm is worse by a factor of two. These observations are no
evidence of superiority of one method over another, but an example
of how to compare and use the different partitioning algorithms on a
given example.

\subsection{Example 2 -- Beyond the resolution limit of modularity: the small
time limit of the continuous process}

Recently, it has been shown that modularity optimization cannot
produce partitions smaller than a certain relative size. This
effect, termed the resolution limit of modularity, leads to
partitions coarser then the expected `natural' community
structure~\cite{FortunatoBarthelemy07}. So far, based on the
discrete-time stability~(\ref{eq:stability}), our analysis has shown
that at time $t=0$, the most stable community structure corresponds
to the trivial partition of each vertex in a community, while the
modularity-optimal community structure corresponds to time $t=1$.
For $t>1$,  the most stable community structures are coarser than
those found by modularity optimization. In order to obtain finer
community structures than modularity (i.e., beyond the resolution
limit), we must consider the stability at times between zero and
one. In fact, this regime can be studied within our framework
through the natural extension to the continuous-time version of
Eq.~(\ref{eq:autocovariance}) obtained through substitution of $M^t$
by $\exp{[(M-I)t]}$, where $I$ is the identity
matrix~\cite{LambiotteDelvenneBarahona}. Keeping linear terms in the
small $t$ expansion of the matrix exponential, we get the following
approximation of the stability for small (continuous) times:
\begin{equation}
\label{eq:continuous_stability} r_{c}(t) \simeq (1-t)\, r(0) + t \,
r(1), \quad 0 \leq t \leq 1.
\end{equation}
Note that this linear interpolation recovers modularity $r(1)$ at
$t=1$ and the totally unclustered graph $r(0)$ at time $t=0$. It
also provides an interpretation in terms of Markov time of the
resolution parameter proposed by Reichardt and
Bornholdt~\cite{ReichardtBornholdt06} and is related to a heuristic
proposed by Arenas \textit{et al.}~\cite{ArenasFernandezGomez07}
consisting of the addition of weighted self-loops to the graph.

\begin{figure}
\begin{center}
\includegraphics[width=8cm]{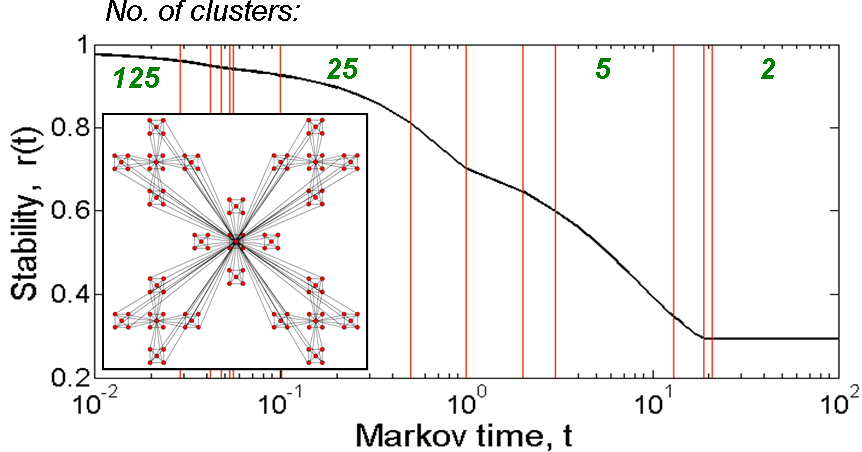}
\end{center}
\caption{Stability curve of a hierarchical, scale-free graph with $N
=125$ vertices proposed in~\cite{RavaszBarabasi03} (shown in the
inset) calculated for times smaller and larger than one. Note that
the natural partitions in 5 and 25 communities have a long time
scale of stability, while the modularity-optimal clustering (at
$t=1$) can be seen as a transient.} \label{FigRavaszBarabasi}
\end{figure}

As an example, Figure~\ref{FigRavaszBarabasi} shows the stability
curve  for times smaller than one of the partitions of a 125-vertex
hierarchical scale-free graph recently proposed by Ravasz and
Barabasi~\cite{RavaszBarabasi03}. In this simple model, the natural
clustering is not found through modularity. Our method, on the other
hand, finds that the natural partitions into 25 and 5 clusters have
long windows of stability while the partition obtained by modularity
at $t=1$ is a transient with no extended significance.
See~\cite{ArenasDiasGuilera07} for another dynamical analysis of the
same graph.

\subsection{Example 3: Structural graphs, model reduction and time scales}

Our final example shows an application of our framework to analyze
graphs of atomic level protein structures and its relevance to model
reduction of biophysical systems. Recently, new methods based on the
explicit consideration of graphs of constraints have been proposed
to simplify the complex dynamics of large biomolecules such as
proteins. The idea is to obtain a simplified, lower-dimensional
mechanical description of the movement of the protein in terms of a
few relatively rigid parts connected by flexible elements
\cite{thorpe,yalibara,hemberg,GfellerDeLosRios08,GfellerDeLosRiosCaflischRao07,reichardtbornholdtprotein}.
Because rigid parts are likely to form a tightly-knit network of
chemical bonds and chemical constraints, while being loosely
interconnected to each other, we expect that a reasonable
approximation to the constrained flexibility of the protein will be
given by the partition of the structural graph of the protein with
atoms as vertices and edges corresponding to bonds and chemical
constraints~\cite{thorpe}.

\begin{figure*}
\subfigure{\includegraphics[width=7.7cm]{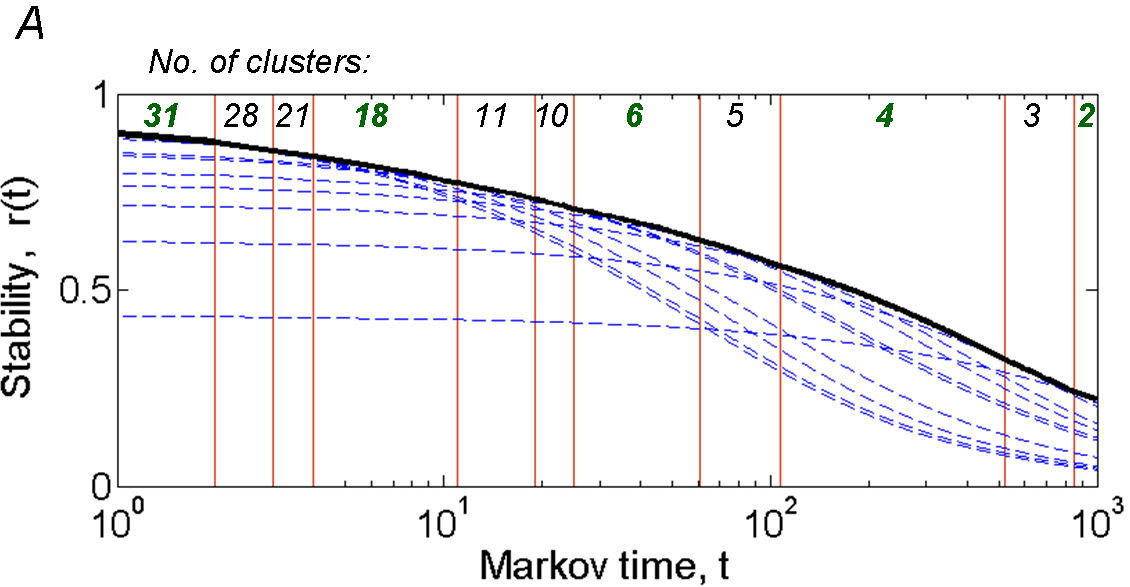}}
\hspace{1cm}
\subfigure{\includegraphics[width=8.1cm]{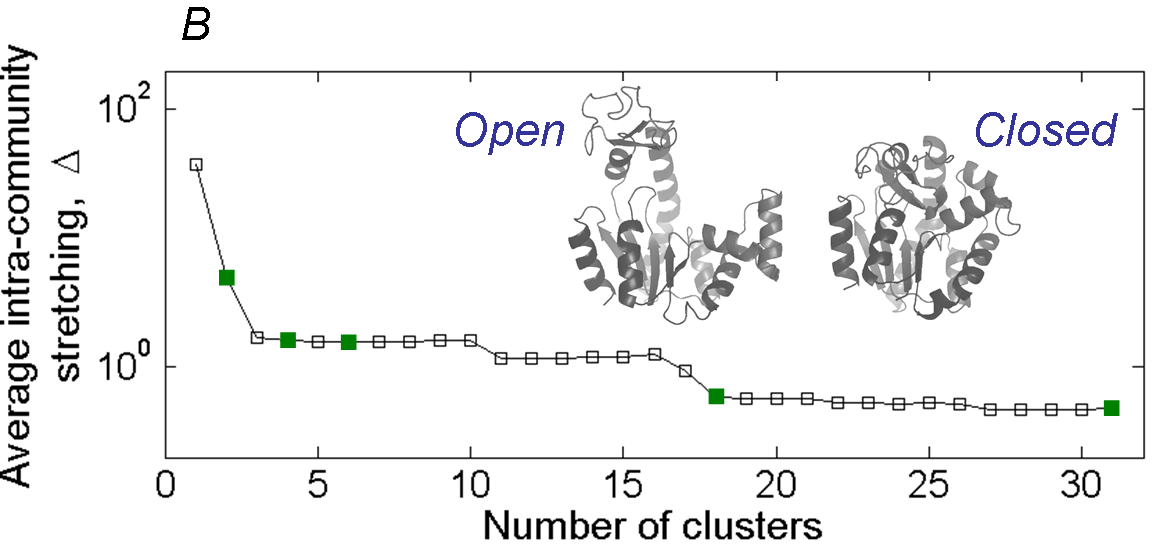}}
\\
\vspace*{-.5cm}
\begin{center}
\subfigure{\includegraphics[width=15cm]{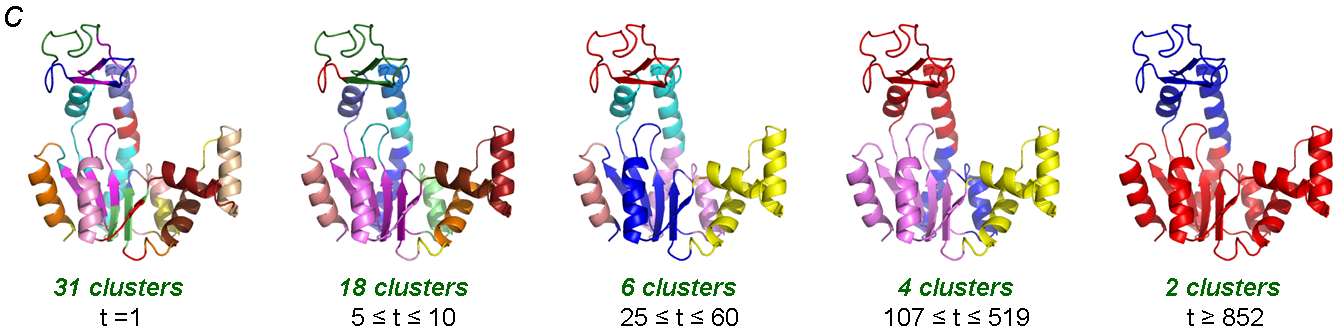}}
\end{center}
\vspace*{-.25cm} \caption{Analysis of the atomic-level structural
graph of the protein Adenylate Kinase (AK) with $N=2085$ vertices.
(See the Supplementary information for a detailed explanation on how
this graph is obtained.) \emph{(A)} The optimal stability curve for
this graph  is estimated by the divisive Shi-Malik algorithm, where
the dashed lines are the stability curves of the different
partitions and the solid curve is the maximum of all dashed curves
at each Markov time. The $31$-way clustering with optimal modularity
among the computed clusterings over-partitions the structure: it
breaks $\beta$-sheets
 and $\alpha$-helices, which should belong to the same cluster.
The $4$-way  and  $18$-way  partitions have relatively long windows
of stability with a good balance between over- and
under-partitioning \emph{(B)} Evaluation of the validity of the
partitions obtained through a comparison of two experimental
conformations of AK (open and closed).  Each partition is obtained
exclusively from the graph of the open configuration. The partitions
are then evaluated against the experimental conformational
distortions to calculate the error obtained by assuming rigidity of
the predicted communities. Two plateaux are observed in the error:
from 4 to 10 clusters and from 18 to 31 clusters. This indicates
that the $4$-way and $18$-way partitions (which show persistence
over long time windows in \textit{(A)}) represent a parsimonious
compromise between rigidity prediction and a small number of
clusters. \emph{(C)} Some of the partitions in the hierarchy of the
system are represented. Note the structural communities (represented
by adjacent regions of the same color) appearing at different Markov
time scales. } \label{protein}
\end{figure*}

Figure~\ref{protein}\textit{A} shows the time hierarchy of
partitions of a full atom ($N=2085$) structural graph of the protein
Adenylate Kinase (AK) in its open configuration. In this example,
biophysical considerations indicate that optimizing modularity
over-partitions the graph---the 31 communities obtained at $t=1$
split several rigid structural motifs such as $\beta$-sheets and
$\alpha$-helices. We use the Shi-Malik divisive algorithm to
estimate the stability curve and obtain a hierarchy of coarser
structures at longer times. Some of the optimal partitions (notably
those into 18 and 4 communities) prevail over relatively long time
windows and contain significant biophysical features. To make this
more precise, we evaluate the relative variation in the
intra-community positions of the $C_\alpha$ carbons of two known
functional configurations of AK (open vs.\ closed) for all
partitions obtained in our study. Figure~\ref{protein}\textit{B}
shows the intra-community stretching for all partitions calculated
as follows: calculate all pair distances between atoms within each
community in both configurations of the protein and obtain $\Delta$,
the average square variation of those distances over all
communities. If the communities are completely rigid, the pair
distances within communities will not change and $\Delta =0$.  The
maximum value $\Delta = 37 \mathring{A}^2$ is the average square
variation for all atoms in the protein (i.e, when we consider all of
them in one community). As the number of communities in the
partition grows, one expects that $\Delta$ will decrease, since the
number of pair distances decreases. The key is to find when the
addition of a community does not result in a significant decrease of
$\Delta$. This implies that the new communities added are not
significantly rigid. This is observed in the plateaux in $\Delta$
that follow the 4-way and 18-way community structures and is
consistent with the extended time scales of prevalence for both
partitions in the stability curve. This indicates that the $4$-way
and $18$-way community structures are a reasonable compromise
between simplicity and predictive power for rigidity. We remark for
this particular example that the `Markov time' is defined as an
abstract entity, not to be assigned an immediate link with a
physical quantity. The rigorous connection between the Markov time
and the biophysical time of protein motions is currently being
pursued.

\section{Discussion and future work}

In this work, we have introduced the stability~(\ref{eq:stability})
as a quality measure of a graph partition. The stability of a
partition is defined in terms of the autocovariance of a Markov
process taking place on the clustered graph and is explicitly
dependent on the Markov time, an intrinsic time scale of the
network. This allows us to rank partitions and establish their
relevance over each time scale.  Although Markov
chains~\cite{vanDongen00,FoussPirotteRendersSaerens06, LatapyPons08}
and dynamical behaviors based on oscillator
dynamics~\cite{ArenasDiasGuileraEtAl06,ArenasDiasGuilera07} have
been used in relation to community detection, previous methods have
not considered the definition of a quality measure, nor have they
introduced the concept of paths of different lengths to evaluate the
quality of partitions across time scales.

The resulting sequence of partitions with maximum stability as a
function of time leads to a time hierarchy of clusterings, from
finer to coarser as the Markov time grows. This hierarchy can be
used to establish the most relevant partitions over the significant
time scales underlying a process. Hence, our method does not provide
a unique partition for the graph. Rather, we propose that, obtaining
the distinct partitions which are valid over different time windows
and selecting those partitions that are relevant over extended time
scales may be better suited for  many applications. In particular,
if a network has been obtained from an underlying dynamical process
with well defined time scales, our analysis can suggest reduced
representations valid over time windows of interest in the process.
On the other hand, if the network under study  does not have an
obvious temporal interpretation, the Markov time acts effectively as
an intrinsic resolution parameter for the partitions.

Another important feature of the stability is that it gives a
unified interpretation in terms of time scales of community
detection methodologies that have been hitherto considered
separately. We have shown that modularity, cut and normalized cut
can be understood  in relation to the stability at $t=1$, while
spectral clustering based on the normalized Fiedler vector is linked
to stability at $t=\infty$. In addition, stability is connected to
the concept of `anti-clustering' and
$k$-colourings~\cite{alon98,ArenasFernFortuGomez08} based on the
existence of recurrence patterns in the time-dependence of the trace
of $R_t$. Although our stability measure~(\ref{eq:stability}) is
defined in the discrete time setting, there is an equivalent
continuous-time version of stability (also introduced above). This
continuous stability can be linked to previous numerical results
where dynamic outcomes, such as synchronization, have been used as
heuristics for graph partitioning~\cite{LambiotteDelvenneBarahona}.
The continuous stability can also be exploited to analyze the regime
beyond the resolution limit of modularity to obtain partitions finer
than those obtained by modularity. In fact, one can show that
previously proposed \textit{ad hoc} multi-resolution
measures~\cite{ReichardtBornholdt06} can be interpreted in terms of
a linearization of the continuous stability at small times.

Complex systems, from protein dynamics to metabolic and social
interactions to the internet, are often described as networks. The
methodology presented here, which extends seamlessly to undirected
weighted graphs, uses the intimate connection between structure and
dynamics to identify communities that can be revealing of the
network structure. In some cases, the original networks are static
and our dynamical approach is a convenient construct to reveal the
intrinsic resolution scales of the problem.  If the network has a
dynamic origin, or indeed it can be related to a Markov
process~\cite{GfellerDeLosRiosCaflischRao07,reichardtbornholdtprotein},
the analysis of the stability of the resulting graph provides
information about the hierarchy of time scales of the underlying
landscape of the system. From this dynamic viewpoint, the presence
of communities relevant over particular time scales hints at a first
step towards reduced representations in which the communities can be
lumped into aggregate variables.  The extension of this methodology
to test systematically for reduced models or model reduction schemes
will be the object of further research.

\subsection{Acknowledgments}
We thank Renaud Lambiotte, Jo\~{a}o Costa and Vincent Blondel for
helpful discussions and Jo\~{a}o Costa for providing the graph and
figures for the AK protein. J.-C. D. and M.B. acknowledge support
from the Mathematics and Life Science Interface panels of the UK
EPSRC.  J.-C. D. holds a FNRS fellowship and is supported by the
Belgian Programme of Interuniversity Attraction Poles and an ARC of
the French Community of Belgium.

%\bibliography{biblio_clustering_2}

\end{document}